\documentclass[epj,final]{svjour}
\usepackage{amsmath}
\usepackage{amssymb}
\usepackage{graphicx}
\usepackage{dcolumn}
\usepackage{bm}

\usepackage{srcltx} 
\begin{document}

\title{Low-energy excitations in the three-dimensional random-field Ising model}


\author{M. Zumsande\inst{1}\inst{2} \and A. K. Hartmann\inst{3}}
\institute{Institute of Theoretical Physics, University of G\"ottingen,
37073 G\"ottingen
\and
Max-Planck Institut fof Physics of Complex Systems,
01187 Dresden, Germany
\and
Institute of Physics, University of Oldenburg, 
26111 Oldenburg, Germany \email{a.hartmann@uni-oldenburg.de}}

\date{\today}

\abstract{The random-field Ising model (RFIM), one of the basic models for
quenched disorder, can be studied numerically with the  help of efficient
ground-state algorithms. In this study, we extend these algorithm by
various methods in order to analyze low-energy excitations for the 
three-dimensional
RFIM with Gaussian distributed disorder that appear in the form of
clusters of connected spins. We analyze several
 properties of these clusters. Our results support
the validity of the droplet-model description for the RFIM.
}


\maketitle

\section{Introduction}
The random-field Ising model (RFIM) \cite{imry1975}
is one of the most basic models
with quenched disorder. Similar to the more prominent spin glasses (SGs)
\cite{mezard1987,fischer1991,young1998}, 
there are still many open questions concerning the
low-temperature properties of the RFIM. 
During the last few years, the RFIM and the related
diluted antiferromagnet in a field have attracted growing attention
\cite{sabhapandit2002,rosas2004,ye2004,colaiori2004,glaser2005,mueller2006,spasojevic2006,lee2006,deAlbuquerque2006,dotsenko2007,silevitch2007,korney2007}, 
in particular within simulation studies at finite \cite{juanjo2003,magni2005,wu2005,prudnikov2005,wu2006,hernandez2008,fytas2008}, and zero
temperature
 \cite{HartmannYoung02,middleton2002,seppala2002,rfim4d2002,dukovski2003,hamasaki2004,alava2005,hamasaki2005,sarjala2006,son2006,santoro2006,liu2007,hastings2008}.

\begin{sloppypar}
There are few results \cite{middleton2002,fes_rfim2002} 
which give evidence that  the low-temperature behavior of 
the three-dimensional RFIM is well described by the 
droplet theory \cite{mcmillan1984,bray1987,fisher1986,fisher1988},
which is one of the most important and most successful theories
to describe finite-dimensional systems exhibiting quenched disorder.
The droplet theory has already turned out to be
 useful to describe the behavior of two-dimensional (2d) SGs 
\cite{stiff2d,droplets2003,droplets_long2004,joerg2006}.
For the 2d SG model, much
evidence supporting the validity of the droplet-model description 
has been accumulated over the years in particular by studying
low-energy excitations. Nevertheless, for the three-dimensional (3d) RFIM, 
low-energy excitations have been investigated only in few cases
\cite{middleton2002,fes_rfim2002} 
so far. In particular, since the 3d RFIM exhibits a phase
transition at non-trivial disorder \cite{bricmont1987}, 
in contrast to the 2d SG,
it is of high interest to study the excitations as a function of the
disorder strength. 
Thus, in this paper we study three different
types of ``typical'' low-energy excitations. Our results show that
the behavior in all three cases is compatible with the droplet
theory, giving strong evidence for the validity of this approach for the RFIM.
 In particular, the different excitations behave the same, for
example concerning their fractal properties. We also show that 
the 
generated excitations exhibit the largest number of spins
close to  the phase transition.
Furthermore, the  distribution of cluster radii is well described
by a power-law $R^{-\theta}$ with $\theta\approx -1.49$ being
the droplet scaling exponent \cite{middleton2002}.
\end{sloppypar}

The RFIM consists of $N$ Ising spins $s_i=\pm 1$ on a regular lattice with 
nearest-neighbor interactions of strength $J$. Additionally, site-dependent
magnetic
fields $h_i$, which are chosen according to some random distribution,
act on each lattice spin. Throughout this paper, a Gaussian
distribution of width $h$ is applied.  Hence, the value of $h$ measures the
strength of the disorder.
The Hamiltonian of the RFIM given by
\begin{equation}
\label{eq:rfim}
H=-J \sum_{\langle ij \rangle} s_i s_j - \sum_{i=1}^N h_i s_i.
\end{equation}
The sum ${\langle ij \rangle}$ runs over nearest neighbors of spins.
We apply periodic boundary condistions in all directions.

The competition between the nearest neighbor interaction and the
tendency for spin $s_i$ to align with its $h_i$ is responsible for the
complexity of the model.
In the RFIM with a three-dimensional lattice, there is a 
2nd order \cite{middleton2002} phase
transition \cite{bricmont1987}
that separates a ferromagnetically ordered phase existing at low
temperature and low disorder from a
disordered phase with average zero magnetization $m= \sum_i s_i$. This
transition is governed by a zero-temperature fixed point. 
From renormalization group arguments it follows
that it is possible \cite{Nattermann98} 
to study the properties of the RFIM also at $T=0$,
i.e.\ by calculating ground states (GSs). It is
convenient that the GS of the RFIM of arbitrary dimension
can be determined in a time that scales polynomially with system size
by effectively algorithms (see next section). 
The equivalent task in spin glasses is 
NP-hard \cite{Barahona82} which implies that no algorithms to solve it
efficiently are know so far.

In this paper, we examine the phase transition in the
three-dimensional RFIM by analyzing low-energy excitations from the
GS via advanced ground-state methods. We first explain
the algorithms we used (Sec.\ 2) before we present the results (Sec.\ 3).
In the final section, we give a summary.

\section{Methods}
\label{sec:methods}

We investigated the excitations in the RFIM via computer
simulations \cite{practical_guide2009} by using
sophisticated optimization algorithms \cite{opt-phys2001}.
We applied three different methods of generating low-energy
excitations. In each of these methods, the first step is to calculate
the GS of a random RFIM realization. In a second step, the
system is perturbed slightly such that
the GS is made a bit unfavorable. How this is done
specifically differs for the three  method. In any case, in
the last step, the GS of the perturbed system is
determined. The resulting configuration is a low-energy excitation
of the original, unperturbed system. The excited state, which 
consists of
one or more clusters of connected spins, can then be compared with the
GS.  

First, to calculate the exact GSs at given randomness $h$, algorithms
\cite{angles-d-auriac1997b,rieger1998,alava2001,opt-phys2001}
from graph theory
\cite{swamy1991,claiborne1990} were applied. To implement them,
some algorithms from the LEDA library \cite{mehlhorn1999} were utilized.
Here the methods are just outlined. More details can be found in the 
literature cited below or in the pedagogical presentation in 
Ref.\ \cite{opt-phys2001}.
For each realization of the disorder, 
given by the values
$\{h_i\}$ of the random fields,
the calculation works by transforming the
system into a network \cite{picard1975}, calculating the maximum flow
in polynomial time 
\cite{traeff1996,tarjan1983,goldberg1988,cherkassky1997,goldberg1998} and
finally obtaining the spin configuration $\{s_i\}$ from the values of
the maximum flow in the network. 
The running time of the latest maximum-flow methods has a peak
near the phase transition and diverges \cite{middleton2002b}
 there like $O(L^{d+1})$.
The first results of applying these algorithms to random-field
systems can be found in Ref.\ \cite{ogielski1986}. In Ref.\
\cite{rfim3-1999} these methods were applied to obtain the exponents 
for the magnetization, the disconnected susceptibility and the
correlation length from GS calculations up to size $L=80$.
The most thorough study of the GSs of the 3d RFIM so far
is presented in Ref.\ \cite{middleton2002}.

Since the algorithms work only with integer values for all parameters,
a value of $J=10000$ was chosen here,  
and all values were rounded to its nearest integer
value. This discreteness is sufficient, as shown in Ref.\ 
\cite{middleton2002}. 
All results are quoted relative to $J$ (or assuming $J\equiv 1$).

Note that in cases where the GS
is degenerate \cite{foot-degenerate},
it is possible to calculate all the GSs in one
sweep \cite{picard1980}, see also
Refs.\ \cite{daff2-1998,bastea1998}. For the
RFIM with a Gaussian distribution of fields, the GS is
non-degenerate, except for a 
two-fold degeneracy at certain values of the randomness, where there
are zero-energy clusters of spins. Thus, 
it is sufficient to calculate just one ground state here.

We are now going to sketch the different excitation
methods that we used. We assume that for a given realization
$\{h_i\}$ of the local fields a GS $\{s_i^0\}$ has been calculated.
\begin{itemize}
\item[I] \emph{Single spin flip method}: 
In this method, a ``central'' spin $s_{i_0}$ is
picked  randomly and \emph{frozen} to an orientation opposite to its GS
orientation $s_{i_0}^0$. 
This is being done by changing the local field $h_i$ to
$h'_i>6 J$ and choosing the sign such that $s'_i$ is aligned opposite to
its GS orientation, e.g.\ $h'_{i_0}=-7Js_{i_0}^0$ 
After recalculating the GS of the perturbed
system, $s'_{i_0}$ is always different from its GS orientation, but
adjacent spins may have flipped as well if it is energetically favorable.
The set of flipped spins will consist always of exactly one connected
cluster of spins.

\item[II] \emph{Random-excitation method}: The system is perturbed by adding a
set of additional fields $\{\delta h_i\}$ 
of strength $\delta$ on top of the original fields
$h_i$. Here it means that each $\delta h_i$ 
is drawn from a uniform distribution
$[-\delta,\delta]$.  The method has been applied earlier 
by Alava and Rieger to
the two-dimensional RFIM \cite{Alava98}, with a uniform distribution
for the random fields as well as for the perturbations.

\item[III] \emph{$\epsilon$-coupling method}: This method, which has been applied
to spin glasses in \cite{HartmannYoung02}, works in a very similar way 
as the
random-excitation method. The system is perturbed by adding an
additional field $\delta h_i$ 
of fixed strength $\epsilon$ to each $h_i$, however
with a site-dependent sign such that the field always acts against the
GS orientation, i.e.\ $h'_i=h_i-\epsilon s_{i}^0$,
lowering the energy of GS configuration.
\end{itemize}

For the second and third method, the calculated GS of the modified
system $\{h'_i\}$ may yield the previous GS $\{s_i^0\}$, in particular if
the strengths $\delta$ and $\epsilon$ are small.
If the strength is large enough,  the excited state will typically exhibit
for both methods 
several clusters of spins flipped with respect to the original GS.

The size of the resulting excited clusters, i.e.\ the number $N_f$ 
of spins exhibiting $s'_i\neq s_i^0$,  can be analyzed in more
than one way. We determine the overlap
\begin{equation}
\label{eq:q}
q=\frac{1}{N} \sum_{i=1}^N s_i^0 s_i'
\end{equation}
 which characterizes the size of the global excitation, also if it
 consists of multiple connected clusters. It is related to the total
 number of flipped spins $N_f$ by
\begin{equation}
1-q=\frac{2 N_f}{N}.
\end{equation} 
In order to analyze the geometry of the clusters of connected spins,
it is convenient to introduce the following three quantities 

\begin{itemize}
\item the volume $V$ is given by the number of spins in a single cluster
of connected spins,
\item  the surface $A$ for
each cluster is given by the number of bonds that
connect a spin of the cluster with another spin that does not belong
to the cluster,
\item the radius of a cluster we define as the root mean-square distance 
between all spins of a cluster, also sometimes called radius of gyration:
\begin{equation}
2 R^2=\sum_{i,j \in {\rm cluster}}
\frac{\vert\vec{r}_i-\vec{r_j}\vert^2}{V^2}\,.
\end{equation}

This means that a single-spin cluster
has radius 0. 
\end{itemize}

\section{Results}
\subsection{Sensitivity of the GS to perturbations}
\label{s:randexres}
In spin glasses, very small variations of parameters such as the
strength of the bonds or an external field can cause excitations that
affect the entire system. This property of disordered systems
resembles \emph{chaos} in systems where a small deviation from initial
conditions can lead to a totally different state of the system at
later times. However, some people prefer to use the term
``hypersensitivity'' \cite{middleton2002} for this non-dynamical
phenomenon.

Small perturbations of this kind have been analyzed in detail in the
context of spin glasses \cite{Fisher88}, \cite{Bray87}.  In the
two-dimensional RFIM, it was found in \cite{Alava98} that a weak form
of chaos is present.

We first applied the  random-excitation method with strength $\delta$
to the 3d RFIM
to investigate how the method is sensitive to the
disorder parameter $h$.

\begin{figure}[h]
\centerline{\includegraphics[width=0.9\columnwidth]{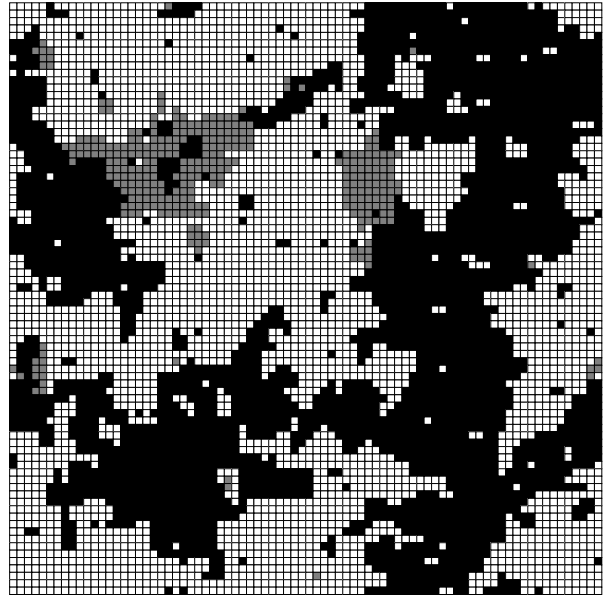}}
\caption{Two-dimensional slice through a 3d RFIM. Spins pointing
$\uparrow$ are marked in white, $\downarrow$-spins in black and spins
that change their orientation in the excitation in Grey.}
\label{g:rand2dexample}
\end{figure}

Most of the resulting excitations consist of flipping clusters at
preexisting interfaces between spins of different orientations, see
Fig.\ \ref{g:rand2dexample}. This is energetically favorable since
only a small number of unsatisfied bonds is created by such an
 excitation. Excitations
inside of domains with ferromagnetic order in the GS are considerably
less frequent. This makes us  expecting
 that close to (or maybe even beyond) the
phase transition, where many domain walls exist, a high number
of excitations is generated.

To gather statistics, we performed simulations for systems with
different $h$ at system sizes ranging from $L=10$ to $L=100$. For
special values $h=2.0$, $h=2.270$, $h=2.40$ and $h=3.0$, we simulated
$n=5000$ samples for each value of $L$, for the remaining values of $h$ the
number of samples is dependent on the system size ($n=1000$ for the
largest systems and a higher number for smaller systems).

We measured  the overlap $q$
as defined in Eq. \ref{eq:q} where $s_i^0$ is the GS
orientation of the spin at site $i$ and $s_i'$ its orientation in the
perturbed state. Note that $q\in[-1,1]$ independent of the system size.

\begin{figure}[h]
\centerline{\includegraphics[width=1.05\columnwidth]{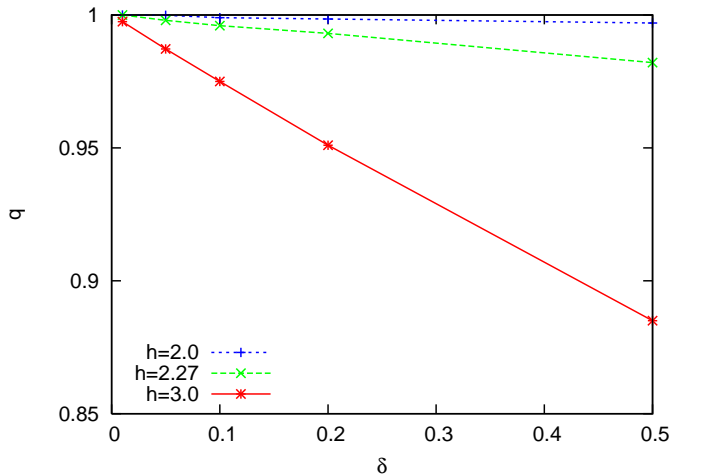}}
\caption{Overlap $q$ of the random excitations (type II) vs.\ 
perturbation strength $\delta$ for various disorder strengths $h$.
\label{g:linresp}}
\end{figure}

We first checked whether the response of the system to the perturbation for
small $\delta$ is linear. The overlap $q(\delta)$ is shown in
Fig.\ \ref{g:linresp}. It was averaged over $5000$ samples of size
$L=40$ for each $(\delta,h)$ pair. The errors are very small, due to
the effect of self averaging, i.e.\ the total overlap does not vary strongly
among different samples.
For different values of $h$,
 and $\delta$ ranging from $\delta=.01$ to $\delta =
0.5$, we found that at least until $\delta=0.2$ the relation between
$q$ and $\delta$ is linear, with a slope that depends on $h$. This
justified to use a fixed $\delta=0.1 J$, as we did throughout our
simulations.

\begin{figure}[h]
\centerline{\includegraphics[width=1.05\columnwidth]{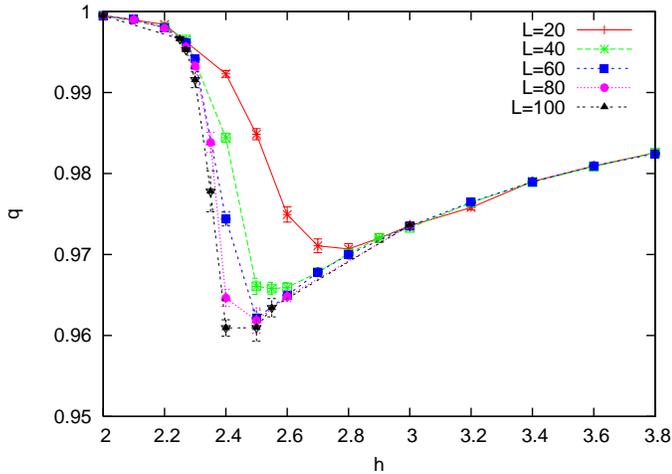}}
\caption{Dependence of the overlap $q(h)$ on the random field strength
$h$ for random excitations (type II) 
of system sizes from $L=20$ to $L=100$. The
perturbation strength is $\delta=0.1J$. The connecting lines are guides
to the eyes.}
\label{g:randexoverlap}
\end{figure}
In Fig.\ \ref{g:randexoverlap}, the overlap $q(h)$ is shown for different
values of the system size $L$. The error bars show the standard error.
For large disorder strength 
$h$, the value of $q$ is independent of the system size and grows slowly
towards $q=1$. In the limit $h \rightarrow \infty$, each spin just follows
its local field independently of its neighbors, which explains this behavior.


Furthermore, one observes that close to the phase transition
point $h_c\approx 2.27$, the overlap is smallest and changes
drastically when starting from small value of $h$.
 The curves appear to be  smooth for small sizes $L$
and significantly steeper at $L=100$.
 Another effect is that with
growing $L$, the minimum of the overlap moves to smaller values of $q$
and closer to the phase transition $h_c\approx 2.27$.
Thus, in the thermodynamic limit $L\to\infty$, one can expect to see
a jump in $q(h)$ when approaching the phase transition
from low values of the disorder $h$.

\begin{sloppypar}
We can compare this behavior with former results of Alava and Rieger
\cite{Alava98} on the two-dimensional RFIM.  For any small fields $h$, the
GS is paramagnetic for the two-dimensional case,
in contrast  three-dimensional RFIM, where the GS is 
ferromagneticaly ordered. 
Yet, the two-dimensional equivalent of
Fig.\ \ref{g:randexoverlap} has a shape similar to the
three-dimensional one with a ``transition'' to $q=1$ for small
$h$. However, this apparent transition is in $d=2$ not an intrinsic property of
the infinite system but a finite-size effect that is caused by the
\emph{breakup length scale} $L_c(h)$. The GSs of
\emph{finite} two-dimensional systems with $L<L_c$ are ferromagnetic
since no domains can exist typically where their random-field energy
exceeds their interface energy. Order is broken only for the infinite system
no matter how small $h$ is, as a consequence of the argument of Imry
and Ma \cite{Imry75}. Therefore, the two-dimensional  transition to
$q=1$ happens at some $h>0$ only in finite systems.
This is reflected by the fact that for
the 2d the apparent transition
point shifts in two dimensions arbitrarily close to $h=0$ with growing
$L$, thus it does not converge to a certain $h_c>0$ as our $d=3$-data
suggests.
\end{sloppypar}

In the cited work of Alava and Rieger, the authors also make
predictions as to what they expect for three dimensions.  In the
limits $h \to 0$ and $h \to \infty$, $q=1$ is expected. However, also
for other $h$, in the thermodynamic limit $q \to 1$ is expected. These
predictions are based on scaling arguments that were developed for
spin glasses \cite{Bray87} and critical exponents from random-bond
models \cite{nattermann1988b} so that it is not clear to what extent
these predictions apply to the RFIM. Our simulations rule out that at
criticality and also at other values of $h$, $q \to 1$, so that only
the predictions in the limit of infinite large and small $h$ are
affirmed.

\begin{figure}[ht]
\includegraphics[width=1.05\columnwidth]{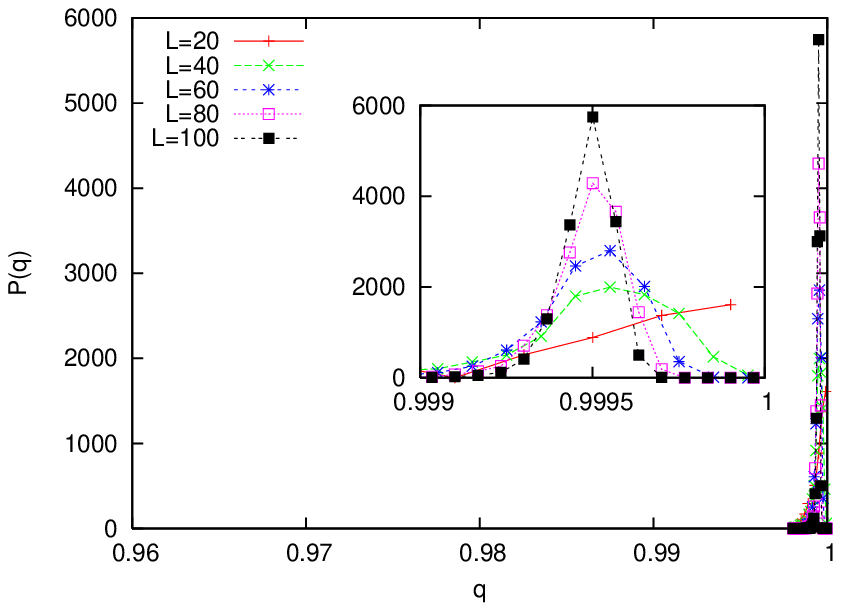}
\includegraphics[width=1.05\columnwidth]{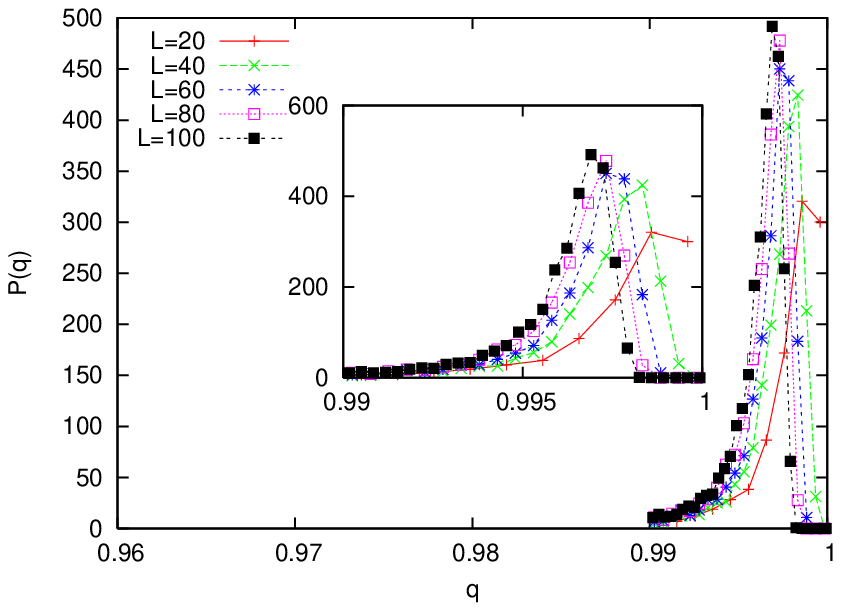} 
\caption{Histograms for the overlap distribution $P(q)$ at $h=2.0$ and
$h=2.2$ for random excitations (type II) with perturbation strength 
$\delta=0.1J$.  The distribution
is sharply peaked very close to $q=1$. Therefore, a zoom to this
region is shown in the respective inset.
Lines are guides to the eyes only, scales are consistent with Fig.~\ref{g:randexhistosB}.}
\label{g:randexhistos}
\end{figure}

\begin{figure}[ht]
\includegraphics[width=1.05\columnwidth]{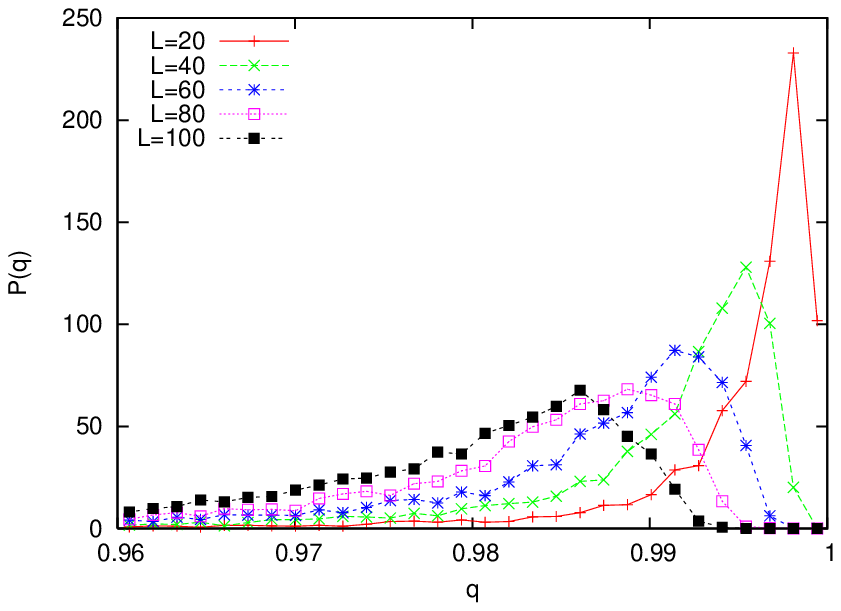}
\includegraphics[width=1.05\columnwidth]{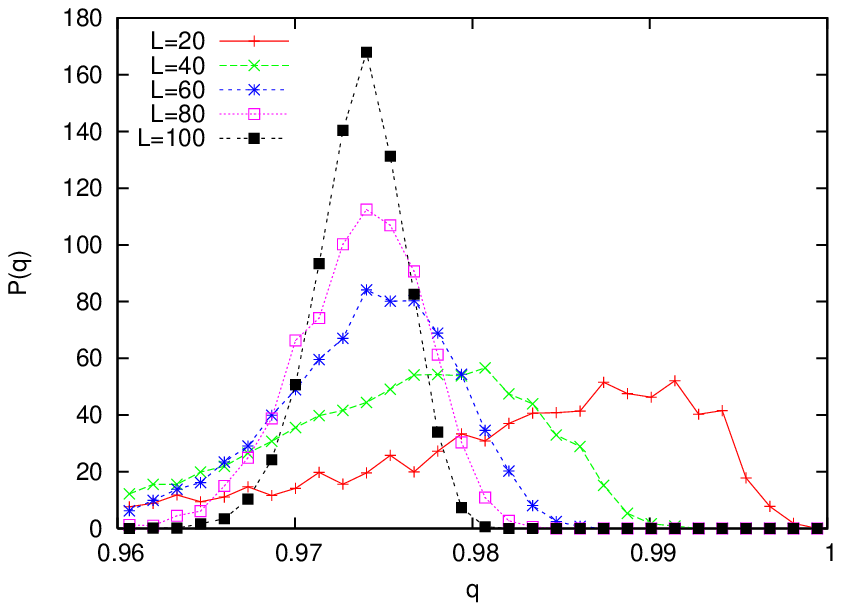}
\caption{Histograms for the overlap distribution $P(q)$ at
 $h=2.4$ and $h=3.0$. Lines
are guides to the eyes only.}
\label{g:randexhistosB}
\end{figure}

For four selected values in the vicinity of $h_c \sim 2.270(4)$
\cite{HartmannYoung02,middleton2002}, 
namely $h=2.0$, $h=2.270$, $h=2.40$ and $h=3.0$,
we analyzed not only the average of the overlap between ground state
and excited state but also the distributions
of the overlaps. The results are shown in Figs.\ \ref{g:randexhistos}.
and \ref{g:randexhistosB}.
In the ferromagnetic phase, represented by $h=2.0$, there is a peak
very close to $q=1$. With increasing system size, the peak becomes
sharper, so that in the thermodynamic limit, $P(q)$ clearly approaches
a $\delta$-shaped peak which is close to $q=0.9995$. The overlap
distribution in the paramagnetic phase, represented by the $h=3.0$
plot, shows a behavior that is in some way similar. For smaller
systems, there is a rather broad distribution with a maximum near
$q=0.975$. The width of the distribution becomes smaller with
increasing system size, and the location of the peak converges to $q
\approx 0.975$.
At $h=2.4$, slightly above $h_c$, there is a transition from the
ferromagnetic peak to the onset of a developing peak at a lower $q$
that will probably become sharper for larger systems. (The maximum of
the $L=80$ distribution is slightly smaller than the maximum of the
$L=100$ distribution.)
At the critical point with $h_c=2.27$, no clear statement can be made:
There is a peak at $q\approx 0.996$ that resembles the ferromagnetic
case, but its width stays approximately constant for the system sizes
we could simulate. However, it is impossible to predict the shape of
the overlap distribution in the thermodynamic limit.

\begin{figure}[h]
\centerline{\includegraphics[width=1.05\columnwidth]{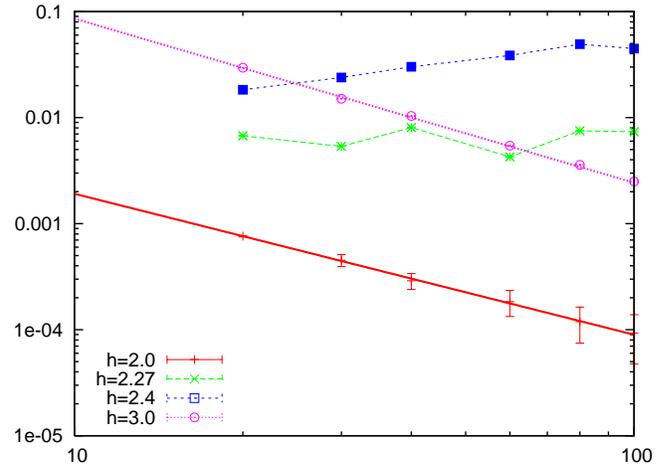}}
\caption{Dependence of the standard deviations $\delta q$ of the
overlap probability distributions of Fig.\ \ref{g:randexhistos} on $L$
($h=2.0,2.27,2.4,3.0$). For $h=2.0$ and $h=3.0$, the lines
represent power laws, while for $h=2.27$ and $h=2.4$, the lines
are guides to the eyes only.}
\label{g:randsigma}
\end{figure}

The standard deviations $\delta q$ of the distributions that are shown
in Fig.\ \ref{g:randsigma} support this picture. While for $h=2.0$ and
$h=3.0$, the width of the distribution decreases with a power law
$\delta q \sim L^{-\alpha}$ with $\alpha=1.24(4)$ ($h=2.0$) and
$\alpha=1.54(3)$ ($h=3.0$), no clear tendency can be seen at the
intermediate $h$. 
If there existed in the RFIM a complex hierarchical phase-space organization
 resembling the replica-symmetry broken phase of the mean-field
SG,   the distribution of overlaps would not
diverge to a $\delta$-peak in that phase. Instead, dependent on the
type of replica symmetry breaking, one would would expect a
distribution that is double-peaked or even flat in the thermodynamic
limit. Although we could not make final predictions for $h_c$, our
results and previous work on the sensitive
of the GS to changes of the boundary conditions \cite{middleton2002} 
suggest that for even larger systems for {\em all} values of $h$ a
$\delta$-peaked distribution should appear.

Note that the distributions we found can be compared with overlap probability
distributions of uncorrelated thermal states of the RFIM in
equilibrium, as performed in \cite{Sinova01}. In that work, the
authors equilibrated samples by MC simulations at a low
temperature. After that, the overlap distribution between the states
in equilibrium was measured. 
The resulting $P(q)$ distributions do resemble each other strongly,
which is in principle a property of a complex phase space.
Nevertheless, the system sizes that could be
equilibrated in Ref. \cite{Sinova01}
were too small ($L<11$) to draw solid conclusions from these results.

\subsection{Fractal dimension of clusters close to $h_c$}
Droplets that represent low-energy excitations in disordered systems
often exhibit a fractal structure. For the RFIM, Middleton and
Fisher \cite{middleton2002} created domain walls by comparing the GS
configurations of different boundary conditions in each sample.  After
calculating the GS where the spins on the left and right border in
$x$-direction are fixed to $\uparrow$-orientation, the GS
was recalculated with the spins on the right border fixed to
$\downarrow$-orientation while the spins on the left border stay fixed
in $\uparrow$-orientation. This method guarantees that a domain wall
is created.

The fractal surface dimension of these domain walls at $h_c$ was
determined to be $d^d_s=2.30 \pm 0.04$. Middleton and Fisher also
analyzed the fractal properties of clusters as areas of equal spin
orientation in the simple pure GS with the result $d^c_s=2.27 \pm
0.02$.

We want to determine whether the fractal dimension of low energy
excitations is compatible to these results. Theory suggests that the
fractal dimensions of excited clusters and domain walls should be
identical if the droplet model applies. This is indeed the case
for 2d Edwards-Anderson spin glasses \cite{HartmannYoung02}. But
for the RFIM, system-wide non-domain-wall excitations that are 
uncommon because
$\theta > 0$, i.e.\ the size of typical droplets is small and not
system-spanning. Note that the fact that typical droplets are small prevents
us from a direct determination of the value of $\theta$ for the droplet
excitations, see next section. This is in contrast
to the domain-wall excitations \cite{middleton2002}, which are
always of the order of the system size. 
This is also in contrast to the 2d SG
model, where $\theta<0$. In this case, droplets tend to be large,
hence the typical length scale is also given by the system size $L$.

We return to the fractal dimension, which
 is determined via measuring the following three quantities,
 see Sec.\ \ref{sec:methods}: We define
the volume $V$ of a droplet as the number of spins it contains. If
there are ``holes'', i.e.\ areas of non-excited spins, inside of an
excitation cluster they do not contribute to $V$. The surface $A$ is
defined as the number of bonds that connect a spin of the cluster with
a spin that is not in the cluster.
For measuring the spatial extension of a cluster, we use the
radius (of gyration).

\begin{figure}[h]
\centerline{\includegraphics[width=1.05\columnwidth]{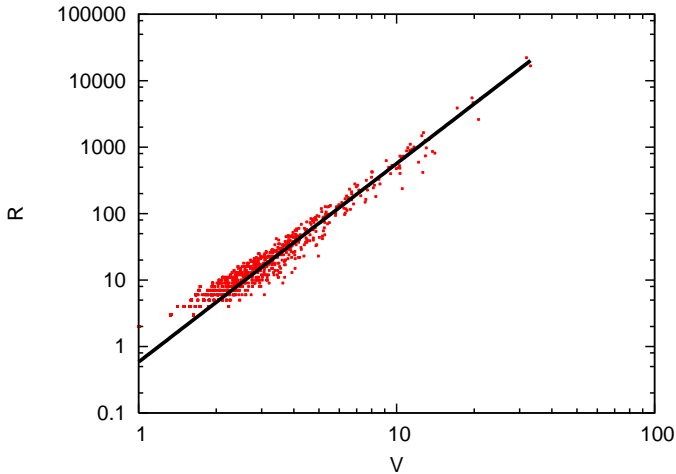}}
\caption{Volume as a function of the 
radius for single spin excitation clusters at
$L=80$. Each data point corresponds to one cluster. A fit to the
function $f(x)=a x^b$ yields $a=0.59(2)$ and $b=2.99(1)$, confirming
$V\sim R^3$, as shown by the line.}
\label{g:midvofr}
\end{figure}

As first result we obtained that the clusters are compact, i.e.\ the volume $V$
is non-fractal: By plotting the volume $V$ as a function
of the radius of gyration $R$ for the one-spin flip method (type I) for
$L=80$ ($h=h_c$), see Fig.\ \ref{g:midvofr}, 
we find a power law with an exponent
of $2.99(1)$, so that $d=3$ fits the data well. Therefore, holes
inside of excited clusters are so rare and small that they do not play
an important role in the excitations of the RFIM. The clusters of the
other methods that were applied to calculate $d_s$ are compact as
well (no plots shown here).

 The surface scales like $A \sim R^{d_s}$ with the surface
dimension $d_s$ that is possibly fractal.  Combined with the
compactness of the clusters, which  means $V \sim R^d$, it follows that
\begin{equation}
A \sim V^{d_s/d}\,.
\end{equation}

\begin{figure}[h]
\centerline{\includegraphics[width=1.05\columnwidth]{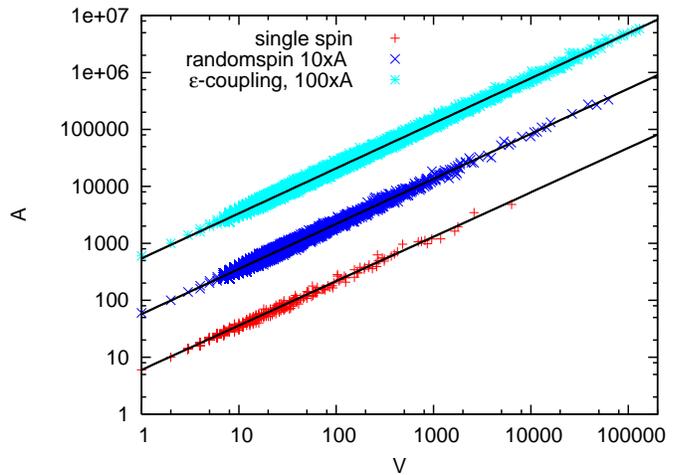}}
\caption{Excitation-cluster surface as a function of the cluster volume. 
 for 
the $\epsilon$-coupling method (top),
the random-excitation method (middle),  
and the single-spin flip method (bottom). 
In all three cases we observe a $A \sim V^{d_s/d}$ behavior (shown
as lines) with $d_s$ being the fractal dimension, as
shown in (\ref{eq:results}).
}
\label{g:vofaall}
\end{figure}

In Fig.\ \ref{g:vofaall}, $A(V)$ is shown for the three different
methods described above. In the double-logarithmic plot, the data of
the random spin method and the $\epsilon$-coupling method is shifted
by multiplying it with a factor of $10$ resp. $100$ in order to make
all three curves visible in one diagram. For the pure data, the curves
overlap of course.  The exponent $d_s/d$ can be extracted from the
numerical data by fitting a power law to  $A(V)$.

From Fig.\ \ref{g:vofaall}, it can be seen that for the three different
methods, the clusters have approximately the same fractal
properties. We fitted a power law of the form $A=c V^{d_s/d}$ to the
data of each method.  The resulting fractal dimensions for each method
are
\begin{equation}
\begin{array}{ll}
d_s=2.34(2) & \mbox{single spin flip} \\ 
d_s=2.37(1) & \mbox{random excitations}\\
d_s=2.36(1) & \mbox{$\epsilon$-coupling} 
\end{array} 
\label{eq:results}
\end{equation}

This means the three different types of excitations behave similar 
within error bars.
Compared with the fractal dimensions calculated in \cite{middleton2002}
($d^c_s=2.27 \pm 0.02$ for clusters and $d^d_s=2.3\pm 0.04$ for domain
walls), our fractal dimension of small excitations is compatible with
the exponent for domain walls $d^d_s$.  The value for the clusters
is a bit off, but this can be expected since these objects are
not covered by the droplet theory.
This agreement, together
with the compactness of all excitation types, verifies one of the
main assumptions of the droplet theory, namely that all types
of physically significant excitations behave the same.

\subsection{Size distributions of clusters at $h_c$}
\label{s:comparesizes}

Another main assumption of the droplet theory \cite{fisher1988} is that
the energy $\Delta E$ of minimum (free-) energy droplets 
with a fixed center and a given scale $l$  
follows a probability distribution
\begin{equation}
P_l(\Delta E) = \frac{1}{l^\theta}\overline{P}(\Delta E/l^\theta)\,,
\label{eq:droplet:scaling}
\end{equation}
where $\theta$ is the {\em droplet-scaling exponent} and $\overline{P}(.)$
is an universal function. Within the droplet theory, as already indicated 
above, it is assumed that the droplet 
exponent $\theta$ describes universally also other types of excitations,
such as system-spanning domain-walls, which can be obtained numerically, 
as described in the previous section.
 Using such an approach, recently the
value of $\theta=1.49(3)$ has been determined for domain walls
\cite{middleton2002}.

Nevertheless, for the RFIM, we are not aware of a direct determination of
the value of $\theta$ via droplet-like excitations. This is in contrast to
the spin-glass case, where for the two-dimensional (2d) systems numerical
simulations  \cite{stiff2d,aspect-ratio2002,excited2d,%
droplets2003,droplets_long2004} 
indicate that domain-wall and droplet-like excitations are
described by one single droplet exponent. For the 2d spin-glass case,
the droplets were generated by a variant of the single-spin-flip method used
in this work. Note that for 2d spin-glasses, the value of $\theta$ is negative,
such that the excitations automatically tend to be as large as possible,
such as to obtain 
an excitation energy as small as possible. Thus, in this case,
the droplet scale $l$ is on average automatically given by the system size $L$.

In case the value of $\theta$ is positive, it is more difficult to determine
its value via the calculation of droplet-like excitations. The reason is that
size of the excitations tends to be small, as mentioned above.
This means, the scale of the excitations is not given by the
system size $L$, in particular each excitation will have its own
scale, compatible with the minimum-energy requirement.
One could use in principle a different approach to 
generate true droplet-like excitations, as
required by the droplet theory, by optimizing only among all clusters
of a given scale $l$, i.e.\ within a range of sizes. Nevertheless, there are
no efficient optimization algorithms available, which can perform this
task, in particular maximum-flow algorithms cannot be applied.
It is quite likely that the problem of miniming
the energy of an excitation under a size-constraint belongs even for the RFIM
to the class of NP-hard
problems. This means that
 only algorithms are known, where the running time increases
in the worst case like an exponential with the system size, limiting 
drastically the size of tractable samples.

Therefore, we follow a different approach here: We want to use
the assumption (\ref{eq:droplet:scaling}) to calculate the properties
of the presently obtained excitations. As a first step, 
we want to calculate the
joint probability $P(R,\Delta E;\, L)$ 
that, for a system size $L$, an minimum-energy excitation
with fixed center  exhibits  the energy $\Delta E$ and 
has scale $R$, here as given by the radius of the excitation
cluster (see Sec.\ \ref{sec:methods}). 
Since the energy of the excitation is minimum, it means that 
$P(R,\Delta E;\,L)$ 
is given by the probability that on (imaginary fixed) scale $R$ the minimum
excitations energy is given by $\Delta E$ and by the probabilities that
for all other scales $l\le L$ with $l\neq R$, the excitation energy is higher.
If we assume that for a fixed scale $l\le L$ the probabilities are independent
of the system size $L$, we obtain:
\begin{equation}
P(R,\Delta E;\,L) = P_R(\Delta E)\prod_{l\neq R; l\le L} 
{\rm Prob}_l(\tilde E >\Delta E)
\label{eq:PLRDE}
\end{equation}
where ${\rm Prob}_l(\tilde E >\Delta E)$ is the probability to obtain,
for a fixed center and a fixed scale $l$ a minimum-energy droplet excitation
larger that $\Delta E$:
\begin{eqnarray}
{\rm Prob}_l(\tilde E >\Delta E) & = & 
1- {\rm Prob}_l(\tilde E \le \Delta E) \nonumber \\
& = & 1- \int_0^{\Delta E} dE' P_l(E') \nonumber\\
 & \stackrel{(\ref{eq:droplet:scaling})}{=} & 
1- \int_0^{\Delta E} dE' \frac{1}{l^\theta}
\overline P(E'/l^\theta) \nonumber \\
& = & 1- 
\int_0^{\Delta E/l^\theta} dx \overline P(x) \nonumber \\
& =: & 1- \overline{Q}
(\Delta E/l^\theta)\,,
\label{eq:cumulative_prob}
\end{eqnarray}
Eq.\ (\ref{eq:PLRDE}) can be rewritten as
\begin{eqnarray}
P(R,\Delta E;\, L) & = & \frac{1}{R^\theta} 
\frac{\overline{P}\left(\frac{\Delta E}{R^\theta}\right)}
{1-\overline{Q}\left(\frac{\Delta E}{R^\theta}\right)}
\prod_{l\le L} 
\left[1- \overline{Q}\left(\frac{\Delta E}{l^\theta}\right)\right]
 \nonumber \\
& = & \frac{1}{R^\theta} 
\frac{\overline{P}\left(\frac{\Delta E}{R^\theta}\right)}
{1-\overline{Q}\left(\frac{\Delta E}{R^\theta}\right)}
e^{ \sum_{l\le L}^\star 
\ln \left[1- \overline{Q}\left(\frac{\Delta E}{l^\theta}\right)\right]}
\label{eq:PRDEL2}
\end{eqnarray}

Note that the sum $\sum_{l\le L}^\star$  is performed according the
assumptions of the droplet theory over different scales. Thus, it
is the same as writing $l=b^k$ for some suitably (basically arbitrarily)
 chosen base $b$ and $\sum_{l\le L} f(l) \longrightarrow \sum_{k=0}^{\log_b L}
f(b^k)$.

Here, using $P(R,\Delta E;\,L)$,
 we are interested in the distribution of excitation radii:
\begin{equation}
P_L(R) := \int_0^{\infty} d\Delta E\,P_L(R,\Delta E;\, L)\,.
\end{equation}
Unfortunately, this integral over $\Delta E$
 cannot be performed analytically. 
Nevertheless, the exponential (third) factor in (\ref{eq:PRDEL2})
 does not depend on $R$.
Furthermore, we can
 assume that the factor $1/R^\theta$ is dominating $P_L(R)$, i.e.\
different contributions from $\Delta E/R^\theta$ arising in the second
factor $\overline P/(1-\overline Q)$ will cancel to first order. 
In other words, the Taylor expansion of the second term yields
a constant plus higher orders in $\Delta E/R^\theta$.  
This is not unreasonable: If we consider the case of the
exponential distribution, where for $\overline P(x)=e^{-x}$ we have
$\overline{Q}(x)=1-e^{-x}$, even all contributions
from $\Delta E/R^\theta$ cancel. In this case, just for completeness,
one finally obtains,
using $\lambda(L) := \sum_{k=0}^{\log_b L} (b^{-\theta})^k$ 
$=(1-(b^{-\theta})^{\log_b L+1})/(1-b^{-\theta})$
$=(1-b^{-\theta}L^{-\theta})/(1-b^{-\theta})$:

\begin{eqnarray}
P_L(R) & = & \frac{1}{R^\theta}\int_0^{\infty} d\Delta E
e^{\Delta E \sum_k^\star (b^\theta)^k} \\
& = & \frac{1}{R^\theta}\int_0^{\infty} d\Delta E
e^{\Delta E \lambda (L)}
= \frac{1}{R^\theta} \frac{1}{\lambda(L)}
\end{eqnarray}
\begin{sloppypar}
Indeed, the probabilities $P_L(R)$ are normalized, since
$\sum_{R\le L}^\star P_L(R)$ $= \lambda(L)\frac{1}{\lambda(L)}$ $= 1$.
\end{sloppypar}

\begin{figure}[ht]
\centerline{\includegraphics[width=0.99\columnwidth]{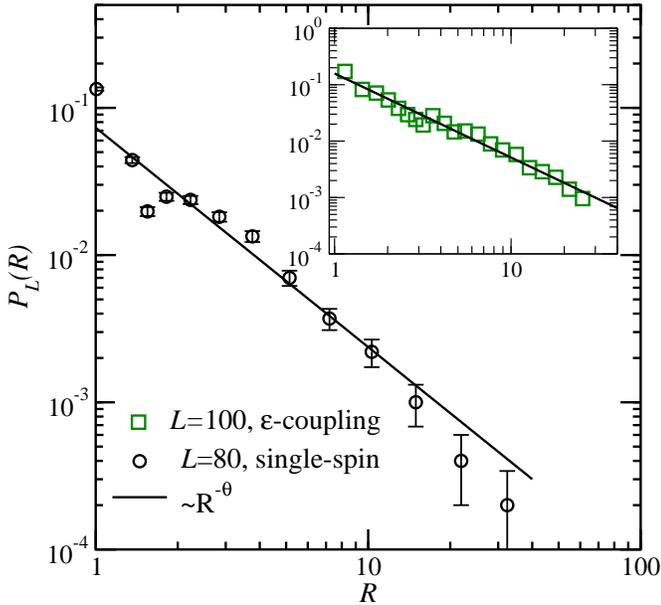}}
\caption{Probabilities $P_L(R)$ of cluster radii $R$ for 
single-spin-flips at $h=h_c$, the inset shows the same quantity for excitations
generated by the $\epsilon$-coupling approach. 
Note that the probabilities are measured (i.e.\ integrated) over
logarithmic scales,
in accordance with the droplet definition within the droplet theory.
The solid lines show 
the power laws $P(R)\sim R^{-\theta}$, which can be 
expected for single-spin-flip excitations (see text).
\label{pic:Pl}
}
\end{figure}

In Fig.\ \ref{pic:Pl}, the measured probabilities $P_L(R)$ for
$R>0$ are shown for  single-spin-flip excitations for system with $L=80$ 
at the transition
point $h=h_c$. The match with the
assumed scaling form $\sim R^{-\theta}$, using 
$\theta=-1.49$ as obtained \cite{middleton2002}
from the scaling of domain-wall energies at $h=h_c$,
 is reasonable. Also for smaller sizes $L<80$, the distribution looks
similar, but they extend only to slightly smaller radii.
Interestingly,
for example for the $\epsilon$-coupling case the distribution shows
the same power-law behavior (see inset), 
although the  $\epsilon$-coupling excitations
are not generated with a ``central'' spin and although each excitation
generates many excitation 
clusters, in contrast to the assumptions used above (for the third type
of excitations, the statistics is not as good right at $h_c$.)

The results show that the behavior of the 3d RFIM excitations, 
where $\theta>0$ right at $h=h_c$,  follow reasonably
well the assumptions of the droplet-scaling theory, similar to the case
of 2d spin glasses, which is an example for a system 
with $\theta<0$, hence simpler to treat. 
This result provides another strong indication that the behavior
of the RFIM at the phase transition
is indeed described to a large extend by the droplet theory.

\section{Summary}
In this paper we studied the properties of low-energy excitations in
the three-dimensional RFIM via  GS calculations and subsequent
generation of GSs for perturbed systems.  By
tracking the difference of the excited state
with respect to the GS of the unperturbed system, we
found that the overlap $q$ undergoes a transition from $q=1$ to a
smaller value that becomes steeper with growing sample size. The
finite-size behavior of the data is compatible with a convergence
of a drastic change of the overlap right at the critical value
$h_c$. This constitutes a clear difference to the 2d
RFIM that was analyzed by Alava and Rieger.

In the distributions $P(q)$, the phase transition is also visible in
the form of a shifting peak. We did not find any clear evidence of an
interval of the disorder parameter $h$ where the distribution 
would reach anything but a peak in the
limit of infinite systems. Close to the transition $h=h_c$ our system
sizes of even
up to $L=100$ are probably too small to reach the scaling regime.
This provides further evidence against a phase
with a complex phase space, similar to replica-symmetry breaking,
in particular right above the transition point.

The geometry of the excitation clusters was found to be compact (volume
to radius) and
fractal (volume to surface). 
Depending on the method by which the excitations were
generated, the fractal dimension $d_s$ is slightly compared to
domain walls, but not statistically significant. 
Furthermore the probabilities of the excitation cluster radii follow
a power-law behavior $~R^{-\theta}$, with $\theta$ being the droplet-scaling
exponent measured previously for domain walls at the phase transition point.
This means that
two main assumptions of the droplet theory, compactness and
universality of the excitations, are verified by our results.

For future work, it would be desirable to test whether the main assumptions of
the droplet theory hold also for the four-dimensional RFIM, where
also some GS results have been obtained previously \cite{rfim4d2002}.
Furthermore it would be desirable to study the dynamics of the
RFIM within the droplet pictures, for example the scaling
of energy barriers with system sizes.

\begin{acknowledgement}
This work was funded by the {\em VolkswagenStiftung} 
(Germany) within the program ``Nachwuchsgruppen an Universit\"aten'',
  by  the  European  Community DYGLAGEMEM program,
and by the ``Gesellschaft f\"ur Wissenschaftliche
Datenverarbeitung'' (GWDG) G\"ottingen by the allocation of computer time.
\end{acknowledgement}

\addcontentsline{toc}{chapter}{\numberline{}Bibliography}
\bibliographystyle{unsrt}
\bibliography{literatur,alex_refs,more_refs}

\begin{thebibliography}{10}

\bibitem{imry1975}
Y.~Imry and S.-K. Ma.
\newblock Random-field instability of the ordered state of continuous symmetry.
\newblock {\em Phys. Rev. Lett.}, 35:1399, 1975.

\bibitem{mezard1987}
M.~M\'ezard, G.~Parisi, and M.A. Virasoro.
\newblock {\em Spin glass theory and beyond}.
\newblock World Scientific, Singapore, 1987.

\bibitem{fischer1991}
K.~H. Fischer and J.~A. Hertz.
\newblock {\em Spin Glasses}.
\newblock Cambridge University Press, Cambridge, 1991.

\bibitem{young1998}
A.~P. Young, editor.
\newblock {\em Spin glasses and random fields}.
\newblock World Scientific, Singapore, 1998.

\bibitem{sabhapandit2002}
S.~Sabhapandit, D.~Dhar, and P.~Shukla.
\newblock Hysteresis in the random-field ising model and bootstrap percolation.
\newblock {\em Phys. Rev. Lett.}, 88(19):197202, Apr 2002.

\bibitem{rosas2004}
A.~Rosas and S.~Coutinho.
\newblock {Random-field Ising model on hierarchical lattices: thermodynamics
  and ground-state critical properties}.
\newblock {\em {Physica A}}, {335}({1-2}):{115--142}, {APR 1} {2004}.

\bibitem{ye2004}
F.~Ye, M.~Matsuda, S.~Katano, H.~Yoshizawa, DP~D.P.~Belanger, E.~T. Seppala,
  J.~A. Fernandez-Baca, and MJ~M.~J.~Alava.
\newblock {Percolation fractal dimension in scattering line shapes of the
  random-field Ising model}.
\newblock {\em {J. Magn. Magn. Mat.}}, {272-76}({Part 2 Sp. Iss.
  SI}):{1298--1299}, {MAY} {2004}.

\bibitem{colaiori2004}
F.~Colaiori, M.~J. Alava, G~.Durin, A.~Magni, and S.~Zapperi.
\newblock {Phase transitions in a disordered system in and out of equilibrium}.
\newblock {\em {Phys. Rev. Lett.}}, {92}({25}), {JUN 25} {2004}.

\bibitem{glaser2005}
A.~Glaser, A.~C. Jones, and P.~M. Duxbury.
\newblock {Domain states in the zero-temperature diluted antiferromagnet in an
  applied field}.
\newblock {\em {Phys. Rev. B}}, {71}({17}), {MAY} {2005}.

\bibitem{mueller2006}
Markus M\"{u}ller and Alessandro Silva.
\newblock Instanton analysis of hysteresis in the three-dimensional
  random-field ising model.
\newblock {\em Phys. Rev. Lett.}, 96(11):117202, 2006.

\bibitem{spasojevic2006}
D.~J. Spasojevic, S.~Janicevic, and M.~Knezevic.
\newblock {Exact results for mean-field zero-temperature random-field Ising
  model}.
\newblock {\em {Europhys. Lett.}}, {76}({5}):{912--918}, {DEC} {2006}.

\bibitem{lee2006}
S.~H. Lee, H.~Jeong, and J.~D. Noh.
\newblock Random field ising model on networks with inhomogeneous connections.
\newblock {\em Phys. Rev. E}, 74(3):031118, 2006.

\bibitem{deAlbuquerque2006}
Douglas~F. de~Albuquerque, I.~P. Fittipaldi, and J.~R. de~Sousa.
\newblock {Absence of tricritical behavior of the random field Ising model in a
  honyecomb lattice}.
\newblock {\em {J. Magn. Mag. Mat.}}, {306}({1}):{92--97}, {NOV 1} {2006}.

\bibitem{dotsenko2007}
V.~S. Dotsenko.
\newblock {On the nature of the phase transition in the three-dimensional
  random field Ising model}.
\newblock {\em {S. Stat. Mech.}}, {SEP} {2007}.

\bibitem{silevitch2007}
D.~M. Silevitch, D.~Bitko, J.~Brooke, S.~Ghosh, G.~Aeppli, and T.~F. Rosenbaum.
\newblock {A ferromagnet in a continuously tunable random field}.
\newblock {\em {Nature}}, {448}({7153}):{567--570}, {AUG 2} {2007}.

\bibitem{korney2007}
L\'{a}szl\'{o} K\"{o}rnyei and Ferenc Igl\'{o}i.
\newblock Geometrical clusters in two-dimensional random-field {I}sing models.
\newblock {\em Phys. Rev. E}, 75(1):011131, 2007.

\bibitem{juanjo2003}
J.~J. Moreno, H.~G. Katzgraber, and AK~A.~K.Hartmann.
\newblock {Finding low-temperature states with parallel tempering, simulated
  annealing and simple Monte Carlo}.
\newblock {\em {Int. J. Mod. Phys. C}}, {14}({3}):{285--302}, {MAR} {2003}.

\bibitem{magni2005}
A.~Magni and V.~Basso.
\newblock {Study of metastable states in the random-field Ising model}.
\newblock {\em {J. Magn. Magn. Mat.}}, {290}({Part 1 Sp. Iss. SI}):{460--463},
  {APR} {2005}.

\bibitem{wu2005}
Y.~Wu and J.~Machta.
\newblock {Ground states and thermal states of the random field Ising model}.
\newblock {\em {Phys. Rev. Lett.}}, {95}({13}), {SEP 23} {2005}.

\bibitem{prudnikov2005}
V.~V. Prudnikov and V.~N. Borodikhin.
\newblock {Monte Carlo simulation of a random-field Ising antiferromagnet}.
\newblock {\em {J. Exp. Theor. Phys.}}, {101}({2}):{294--298}, {2005}.

\bibitem{wu2006}
Y.~Wu and J.~Machta.
\newblock {Numerical study of the three-dimensional random-field Ising model at
  zero and positive temperature}.
\newblock {\em {Phys. Rev. B}}, {74}({6}), {AUG} {2006}.

\bibitem{hernandez2008}
L.~Hernández and H.~Ceva.
\newblock Wang-landau study of the critical behavior of the bimodal 3d random
  field ising model.
\newblock {\em Physica A}, 387(12):2793 -- 2801, 2008.

\bibitem{fytas2008}
N.~G. Fytas and A.~Malakis.
\newblock {Phase diagram of the 3D bimodal random-field Ising model}.
\newblock {\em {Eur. Phys. J. B}}, {61}({1}):{111--120}, {JAN} {2008}.

\bibitem{HartmannYoung02}
{A.K. Hartmann, A.P. Young}.
\newblock {Large-Scale, Low-Energy Excitations in the Two-Dimensional Ising
  Spin Glass}.
\newblock {\em Phys. Rev. B}, 66:094419, 2002.

\bibitem{middleton2002}
A.~Alan Middleton and Daniel~S. Fisher.
\newblock Three-dimensional random-field ising magnet: Interfaces, scaling, and
  the nature of states.
\newblock {\em Phys. Rev. B}, 65(13):134411, Mar 2002.

\bibitem{seppala2002}
E.~T. Sepp\"al\"a, A.~M. Pulkkinen, and M.~J. Alava.
\newblock Percolation in three-dimensional random field ising magnets.
\newblock {\em Phys. Rev. B}, 66(14):144403, Oct 2002.

\bibitem{rfim4d2002}
Alexander~K. Hartmann.
\newblock Critical exponents of four-dimensional random-field ising systems.
\newblock {\em Phys. Rev. B}, 65(17):174427, May 2002.

\bibitem{dukovski2003}
I.~Dukovski and J.~Machta.
\newblock Ground-state numerical study of the three-dimensional random-field
  {I}sing model.
\newblock {\em Phys. Rev. B}, 67(1):014413, Jan 2003.

\bibitem{hamasaki2004}
T.~Hamasaki and H.~Nishimori.
\newblock {Exact ground-state energies of the random-field Ising chain and
  ladder}.
\newblock {\em {J. Phys. Soc. Japan}}, {73}({6}):{1490--1495}, {JUN} {2004}.

\bibitem{alava2005}
M.~J. Alava, V.~Basso, F.~Colaiori, L.~Dante, G.~Durin, A.~Magni, and
  S.~Zapperi.
\newblock {Ground-state optimization and hysteretic demagnetization: The
  random-field Ising model}.
\newblock {\em {Phys. Rev. B}}, {71}({6}), {FEB} {2005}.

\bibitem{hamasaki2005}
T.~Hamasaki and H.~Nishimori.
\newblock {Exact ground-state energies of the random-field Ising chain and
  ladder}.
\newblock {\em {Progr. Theor. Phys. Suppl.}}, {157}:{120--123}, {2005}.

\bibitem{sarjala2006}
M.~Sarjala, V.~Petaja, and M.~Alava.
\newblock {Optimization in random field Ising models by quantum annealing}.
\newblock {\em {J. Stat. Mech.}}, {JAN} {2006}.

\bibitem{son2006}
S.~W. Son, H.~Jeong, and J.~D. Noh.
\newblock {Random field Ising model and community structure in complex
  networks}.
\newblock {\em {Eur. Phys. J. B}}, {50}({3}):{431--437}, {APR} {2006}.

\bibitem{santoro2006}
G.~E. Santoro and E.~Tosatti.
\newblock {Optimization using quantum mechanics: quantum annealing through
  adiabatic evolution}.
\newblock {\em {J. Phys. A}}, {39}({36}):{R393--R431}, {SEP 8} {2006}.

\bibitem{liu2007}
Yang Liu and Karin~A. Dahmen.
\newblock {No-passing rule in the ground state evolution of the random-field
  Ising model}.
\newblock {\em {Phys. Rev. E}}, {76}({3, Part 1}), {SEP} {2007}.

\bibitem{hastings2008}
M.~B. Hastings.
\newblock Inference from matrix products: A heuristic spin-glass algorithm.
\newblock {\em Physical Review Letters}, 101(16):167206, 2008.

\bibitem{fes_rfim2002}
M.~Zumsande, M.~J. Alava, and A.~K. Hartmann.
\newblock First excitations in two- and three-dimensional random-field ising
  systems.
\newblock {\em J. Stat. Mech.}, page P02012, 2008.

\bibitem{mcmillan1984}
W.~L. McMillan.
\newblock Scaling theory of {I}sing spin glasses.
\newblock {\em J. Phys. C}, 17:3179, 1984.

\bibitem{bray1987}
A.~J. Bray and M.~A. Moore.
\newblock Scaling theory of the ordered phase of spin glasses.
\newblock In J.~L. van Hemmen and I.~Morgenstern, editors, {\em Heidelberg
  Colloquium on Glassy Dynamics}, page 121. Springer, Berlin, 1987.

\bibitem{fisher1986}
D.~S. Fisher and D.~A. Huse.
\newblock Ordered phase of short-range {I}sing spin-glasses.
\newblock {\em Phys. Rev. Lett.}, 56:1601, 1986.

\bibitem{fisher1988}
D.~S. Fisher and D.~A. Huse.
\newblock Equilibrium behavior of the spin-glass ordered phase.
\newblock {\em Phys. Rev. B}, 38:386, 1988.

\bibitem{stiff2d}
A.~K. Hartmann and A.~P. Young.
\newblock Lower critical dimension of {I}sing spin glasses.
\newblock {\em Phys. Rev B}, 64:180404, 2001.

\bibitem{droplets2003}
A.~K. Hartmann and M.~A. Moore.
\newblock Corrections to scaling are large for droplets in two-dimensional spin
  glasses.
\newblock {\em Phys. Rev. Lett.}, 90:12720, 2003.

\bibitem{droplets_long2004}
A.~K. Hartmann and M.~A. Moore.
\newblock Generating droplets in two-dimensional {I}sing spin glasses by using
  matching algorithms.
\newblock {\em Phys. Rev. B}, 69:104409, 2004.

\bibitem{joerg2006}
T.~J\"org, J.~Lukic, E.~Marinari, and O.~C. Martin.
\newblock Strong universality and algebraic scaling in two-dimensional {I}sing
  spin glasses.
\newblock {\em Phys. Rev. Lett.}, 96:237205, 2006.

\bibitem{bricmont1987}
J.~Bricomont and A.~Kupiainen.
\newblock Lower critical dimension of the random-fiel {I}sing model.
\newblock {\em Phys. Rev. Lett.}, 59:1829, 1987.

\bibitem{Nattermann98}
{T. Nattermann}.
\newblock {Theory of the Random Field Ising Model}.
\newblock {\em Spin glasses and Random Fields (ed. P. Young)}, pages 277--298,
  1998.

\bibitem{Barahona82}
{F. Barahona}.
\newblock {On the computational complexity of Ising spin glass models}.
\newblock {\em J. Phys. A}, 15:3241--3253, 1982.

\bibitem{practical_guide2009}
A.~K. Hartmann.
\newblock {\em {Practical Guide to Computer Simulations}}.
\newblock Word Scientific, Singapore, 2009.

\bibitem{opt-phys2001}
A.~K. Hartmann and H.~Rieger.
\newblock {\em Optimization Algorithms in Physics}.
\newblock Wiley-VCH, Weinheim, 2001.

\bibitem{angles-d-auriac1997b}
M.~Preissmann J.~C. Angl\`es~d'Auriac and A.~Sebo.
\newblock Optimal cuts in graphs and statistical mechanics.
\newblock {\em J. Math. and Comp. Model.}, 26:1, 1997.

\bibitem{rieger1998}
H.~Rieger.
\newblock Ground state properties of frustrated systems.
\newblock In J.~Kertesz and I.~Kondor, editors, {\em Advances in Computer
  Simulation, Lecture Notes in Physics}, volume 501, Heidelberg, 1998.
  Springer.

\bibitem{alava2001}
C.~Moukarzel M.~J.~Alava, P. M.~Duxbury and H.~Rieger.
\newblock Combinatorial optimization and disordered systems.
\newblock In C.~Domb and J.L. Lebowitz, editors, {\em Phase transitions and
  Critical Phenomena}, volume~18. Academic press, New York, 2001.

\bibitem{swamy1991}
M.N.S. Swamy and K.~Thulasiraman.
\newblock {\em Graphs, Networks and Algorithms}.
\newblock Wiley, New York, 1991.

\bibitem{claiborne1990}
J.D. Claiborne.
\newblock {\em Mathematical Preliminaries for Computer Networking}.
\newblock Wiley, New York, 1990.

\bibitem{mehlhorn1999}
K.~Mehlhorn and St. N\"aher.
\newblock {\em The LEDA Platform of Combinatorial and Geometric Computing}.
\newblock Cambridge University Press, Cambridge, 1999.

\bibitem{picard1975}
J.-C. Picard and H.D. Ratliff.
\newblock Minimum cuts and related problems.
\newblock {\em Networks}, 5:357, 1975.

\bibitem{traeff1996}
J.L. Tr\"aff.
\newblock A heuristic for blocking flow algorithms.
\newblock {\em Eur.\ J.\ Oper.\ Res.}, 89:564, 1996.

\bibitem{tarjan1983}
R.E. Tarjan.
\newblock {\em Data Structures and Network Algorithms}.
\newblock Society for Industrial and Applied Mathematics, Philadelphia, 1983.

\bibitem{goldberg1988}
A.~V. Goldberg and R.~E. Tarjan.
\newblock A new approach to the maximum-flow problem.
\newblock {\em J. ACM}, 35:921, 1988.

\bibitem{cherkassky1997}
B.~Cherkassky and A.~Goldberg.
\newblock On implementing the push-relabel method for the maximum flow problem.
\newblock {\em Algorithmica}, 19:390, 1997.

\bibitem{goldberg1998}
A.~V. Goldberg and R.~Satish.
\newblock Beyond the flow decomposition barrier.
\newblock {\em J. ACM}, 45:783, 1998.

\bibitem{middleton2002b}
A.~A. Middleton.
\newblock Critical slowing down in polynomial time algorithms.
\newblock {\em Phys. Rev. Lett.}, 88:017202, 2002.

\bibitem{ogielski1986}
A.~T. Ogielski.
\newblock Integer optimization and zero-temperature fixed point in {I}sing
  random-field systems.
\newblock {\em Phys. Rev. Lett.}, 57:1251, 1986.

\bibitem{rfim3-1999}
A.~K. Hartmann and U.~Nowak.
\newblock Universality in three dimensional random field systems.
\newblock {\em Eur. Phys. J. B}, 7:105, 1999.

\bibitem{foot-degenerate}
The RFIM with a delta-distribution of the random fields ($\pm h$) exhibits an
  exponential ground-state degeneracy.

\bibitem{picard1980}
J.-C. Picard and M.~Queyranne.
\newblock On the structure of all minimum cuts in a network and applications.
\newblock {\em Math.~Prog.~Study}, 13:8, 1980.

\bibitem{daff2-1998}
A.~K. Hartmann.
\newblock Ground state structure of diluted antiferromagnets and random field
  systems,.
\newblock {\em Physica A}, 248:1, 1998.

\bibitem{bastea1998}
S.~Bastea and P.~M. Duxbury.
\newblock Ground state structure of random magnets.
\newblock {\em Phys. Rev. E}, 58:4261, 1998.

\bibitem{Alava98}
{M. Alava, H. Rieger}.
\newblock {Chaos in the random field Ising model}.
\newblock {\em Phys. Rev. E}, 58:4284, 1998.

\bibitem{Fisher88}
{D. Fisher, A. Huse}.
\newblock {Nonequilibrium dynamics of spin glasses}.
\newblock {\em Phys. Rev. B}, 38:373--385, 1988.

\bibitem{Bray87}
{A. J. Bray, M.A. Moore}.
\newblock {Chaotic Nature of the Spin-Glass Phase }.
\newblock {\em Phys. Rev. Lett.}, 58:57--60, 1987.

\bibitem{Imry75}
{Y. Imry, S. Ma}.
\newblock {Random-Field Instability of the Ordered State of Continous
  Symmetry}.
\newblock {\em Phys. Rev. Lett.}, 35:1399--1401, 1975.

\bibitem{nattermann1988b}
T.~Nattermann.
\newblock {Ground-state instability of interfaces in random-systems}.
\newblock {\em {Phys. Rev. Lett.}}, {60}({25}):{2701}, {jun 20} {1988}.

\bibitem{Sinova01}
{J. Sinova and G. Canright}.
\newblock {Nature and number of distinct phases in the random-field Ising
  model}.
\newblock {\em Phys. Rev. B}, 64:094402, 2001.

\bibitem{aspect-ratio2002}
A.~K. Hartmann, A.~J. Bray, A.~C. Carter, M.~A. Moore, and A.~P. Young.
\newblock The stiffness exponent of two-dimensional {I}sing spin glasses for
  non-periodic boundary conditions using aspect-ratio scaling.
\newblock {\em Phys. Rev. B}, 66:224401, 2002.

\bibitem{excited2d}
A.~K. Hartmann and A.~P. Young.
\newblock Large-scale, low-energy excitations in the two-dimensional {I}sing
  spin glass.
\newblock {\em Phys. Rev B}, 65:094419, 2002.

\end{thebibliography}

\end{document}